\documentstyle[preprint,aps]{revtex}
\begin{document}
%\draft
\title{\LARGE \bf LATTICE COMPUTATION OF THE EFFECTIVE   \\
\vspace*{3mm}
            POTENTIAL IN O(2)-INVARIANT $\lambda\Phi^4$ THEORY }

\author{ A. Agodi, G. Andronico and M. Consoli}

\address{ Dipartimento di Fisica - Universita di Catania; \\
 Istituto Nazionale di Fisica Nucleare - Sezione di Catania;\\
Corso Italia, 57 - I 95129 Catania - Italy}
\newcommand{\beq}{\begin{equation}}
\newcommand{\eeq}{\end{equation}}

\maketitle

\begin{abstract}
We present a lattice computation of the effective potential
for O(2)-invariant $(\lambda\Phi^4)_4$ theory in the region of bare
parameters
corresponding to a classically scale-invariant theory.
 As expected from ``triviality'' and
as in the one-component theory, we find very good agreement with the one-loop
prediction, while
 a perturbative leading-log
improvement of the effective potential fails to reproduce the Monte
Carlo data. The mass $m_h$
of the free shifted radial field is related to the
renormalized vacuum expectation value $v_R$
through the same relation $m^2_h=8\pi^2 v^2_R$ as in the one-component case.
This confirms the prediction of
 a weakly interacting 2.2 TeV Higgs particle in the standard model.
\end{abstract}
\vskip 15 pt
\widetext

In the study of Spontaneous Symmetry Breaking (SSB) the simplest quantity to
compute is the vacuum expectation value of the bare scalar field $\Phi_B(x)$
in the presence of an external constant source $J$
\beq
      \langle \Phi_B \rangle _J= \phi_B(J)
\eeq
Computing $\phi_B$ on the lattice
at several $J$-values is equivalent \cite{huang} to inverting the relation
\beq
        J=J(\phi_B)={{dV_{eff}}\over{d\phi_B}}
\eeq
involving the effective potential $V_{eff}(\phi_B)$.
In this framework, the occurrence of SSB is determined
from exploring (for $J\neq0$) the properties of the function
\beq
               \phi_B(J)=-\phi_B(-J)
\eeq
in connection with the limiting behaviour at zero external source
\beq
  \lim_{J\to 0^{\pm}}~\phi_B(J)=\pm v_B \neq 0
\eeq
over a suitable range of the bare parameters
$(r_o,\lambda_o)$ in the lattice action
\beq
 a^4 \sum _x~[ {{ \sum^4_{\mu=1}( \Phi_B(x+ae_{\mu})- \Phi_B(x))^2}
\over{2a^2}}
 +
{{1}\over{2}} r_o \Phi_B^2(x) + \frac{\lambda_o}{4} \Phi_B^4(x)]
\eeq
A lattice simulation of the weakly coupled
 one-component massless $\lambda\Phi^4$
theory \cite{ago} has shown that
there is a well defined region in the bare parameter plane $(r_o,\lambda_o)$,
corresponding to SSB from the classically scale-invariant case, where the
effective potential is reproduced by its one-loop form to very high accuracy.
Even though ${{\lambda_o}\over{\pi^2}}<<1$,
 a ``leading-log improvement'' of the one-loop potential
completely fails to reproduce the Monte Carlo data. This result, while
contradicting the naive perturbative expectations, confirms the crucial insight
of \cite{consteve,zeit} on the basis of the generally accepted ``triviality''
\cite{froh} of
$(\lambda\Phi^4)_4$: for a ``trivial'' theory the one-loop potential is
effectively exact in the continuum limit.
 In fact, since there are no observable
interactions, the effective
potential is just the sum of the classical potential and the zero-point energy
of the free fluctuation field $h(x)=\Phi_B(x)-\langle\Phi_B\rangle$.
 The traditional perturbative renormalization, based on the concept
of a cutoff-independent and {\it non-vanishing} renormalized coupling at
non-zero
external momenta $\lambda_R$, is not appropriate \cite{trivpert}
just because it would spoil
this exactness-it does not properly re-absorb infinities but merely pushes
them into ``higher orders'' which are then neglected. In this sense, the
unphysical features of the perturbative renormalization (the one-loop Landau
pole or the spurious two-loop ultraviolet fixed point at non zero bare coupling
which contradicts the rigorous arguments of \cite{froh}) are just a signal of
the inadequacy of the procedure. However, it is simple to renormalize the
one-loop potential exactly \cite{consteve,zeit,bran,con,iban,new,rit}
(this was first discovered
in the context of the Gaussian effective potential \cite{cian,return,cast}).
 The particle mass $m_h$ is related to the ultraviolet cutoff and to the bare
vacuum field $v_B$ through
\beq
      m^2_h=3\lambda_ov^2_B=
\Lambda^2~\exp (-{{16\pi^2}\over{9\lambda_o}}).
\eeq
[Note that $\lambda_o=\lambda_B/6$ in the notation of \cite{consteve,zeit}].
At the same time, the vacuum energy density (a renormalization-group-invariant
quantity) is
\beq
             V^{\rm 1-loop}(\pm v_B)=-{{m^4_h}\over{128\pi^2}}
\eeq
Hence, to get a cutoff-independent $m_h$ in the continuum
limit $\Lambda \to \infty$,
$\lambda_o=\lambda_o(\Lambda)$ has to vanish as $1/(\ln{{\Lambda}\over{m_h}})$
and the bare vacuum field $\phi_B$ is non-trivially rescaled with respect to
the
physical field $\phi_R$ through
\beq
             \phi^2_B=Z_{\phi}\phi^2_R
\eeq
with  $Z_{\phi}\sim 1/\lambda_o$.
The non-perturbative nature of the vacuum field
renormalization ($Z_{\phi}\sim 1/\lambda_o$), first discovered in the
gaussian approximation by Stevenson and Tarrach \cite{return}, should not be
confused with the $h$-field wave function renormalization.
Since $h$ is just a free field,
 one has trivially $Z_h=1$. The structure
$Z_{\phi}\neq Z_h$ is allowed because for a scalar field the
decomposition into $p_{\mu}=0$ and $p_{\mu}\neq 0$ components is
Lorentz-invariant
\cite{consteve,zeit,rit}.
This structure is more
general than in perturbation theory and is the essential ingredient that
allows SSB to coexist with ``triviality''.
 Finally, the physical normalization condition \cite{con,new,consteve,zeit}
\beq
   {{d^2V^{\rm 1-loop}}\over{d\phi^2_R}}|_{\phi_R=\pm v_R}=m^2_h
\eeq
determines
\beq
                    Z_{\phi}={{8\pi^2}\over{3\lambda_o}}
\eeq
and leads to
\beq
                    m^2_h=8\pi^2v^2_R
\eeq
In the context of the standard model,
where $v_R$ is known from the Fermi constant to be 246 GeV, this
predicts a Higgs
mass $\sim$2.2 TeV (up to radiative corrections that are small
if the top mass is below
200 GeV \cite{con,new}).
\par As
discussed in \cite{con,new,consteve}, one expects Eq.(11), obtained in the
single-component theory,
 to be also valid in the O(N)-continuous symmetry case. This observation
originates in Ref.\cite{dj} which obtained
 the same effective potential for the
radial field as in the
one-component theory. This is extremely
intuitive. The Goldstone-boson fields do not contribute non-trivially to the
effective potential: they contribute only their zero-point energy, which is
a constant since these are free, massless fields, according to ``triviality''.
Thus, in the O(2) case, one may take
 the diagram $(V_{eff},\phi_B)$ for the one-component theory
and ``rotate'' it around the $V_{eff}$ symmetry axis. This generates a three-
dimensional diagram $(V_{eff},\phi_1,\phi_2)$ where $V_{eff}$ depends on the
bare radial field,
\beq
                \rho_B=\sqrt{\phi^2_1+\phi^2_2}
\eeq
in exactly the same way as $V_{eff}$ depends on $\phi_B$ in the one-component
theory; namely
($\omega^2(\rho_B)=3\lambda_o\rho^2_B$)
\beq
 V^{{\rm 1-loop}}(\rho_B)  =  \frac{\lambda_o}{4} \rho^4_B +
\frac{\omega^4{\scriptstyle (\rho_B)}}{64\pi^2}
\left( \ln \frac{\omega^2{\scriptstyle (\rho_B)} }{\Lambda^2} -
\frac{1}{2} \right) .
\eeq
This represents the classical potential plus the zero-point energy of the
{\it free} shifted radial
 field. ``Triviality'' implies that there are no observable interaction
effects, so this result should be exact \cite{consteve,zeit}. By using
eqs.(6,10),
 $V^{\rm 1-loop}$
 can be re-expressed in the form
\beq
V^{\rm 1-loop}(\rho_B)={{\pi^2\rho^4_B}\over{Z^2_{\phi} }}
(\ln{{\rho^2_B}\over{v^2_B}}-{{1}\over{2}}) ,
\eeq
\par Differentiating Eq.(14), we obtain the bare ``radial source''
\beq
J(\rho_B)={{dV^{\rm 1-loop}(\rho_B)}\over{d\rho_B}}=
{{4\pi^2\rho^3_B}\over{Z^2_{\phi} }}\ln{{\rho^2_B}\over{v^2_B}} ,
\eeq
which we shall compare
 with the lattice results for $J=J(\rho_B)$.
\par The lattice simulation of the O(2)-invariant theory is obtained from the
action
\[
 a^4 \sum _x~[
{{ \sum^4_{\mu=1}( \Phi_1(x+ae_{\mu})- \Phi_1(x))^2}
\over{2a^2}} +
{{ \sum^4_{\mu=1}( \Phi_2(x+ae_{\mu})- \Phi_2(x))^2}
\over{2a^2}}
\]
\beq
 +{{1}\over{2}} r_o (\Phi_1^2(x)+\Phi^2_2(x))
 + \frac{\lambda_o}{4} (\Phi^2_1(x)+\Phi^2_2(x))^2]
\eeq
and one couples $\Phi_1$ and $\Phi_2$ to two constant external sources $J_1$
and
$J_2$ through
\beq
        a^4\sum_x~[J_1\Phi_1(x)+J_2\Phi_2(x)]
\eeq
By using $J_1=J\cos\theta$ and $J_2=J\sin\theta$ it is straightforward to show
that the bare radial field
($\phi_{1}=\langle\Phi_{1}\rangle_{J_1,J_2}$,
$\phi_{2}=\langle\Phi_{2}\rangle_{J_1,J_2}$)
\beq
          \rho_B=\sqrt{\phi^2_1+\phi^2_2}
\eeq
does only depend on $J$, that is
\beq
\rho_B=\rho_B(J)
\eeq
We started our Monte Carlo simulation on a $10^4$ lattice by investigating
first
the $(r_o,\lambda_o)$ correlation
which corresponds to the classically scale-invariant case.
Analytically, this corresponds to determine $r_o$ from
the zero-mass renormalization
condition \cite{cw}
\beq
{{d^2V_{eff}}\over{d\rho^2_B}}|_{\rho_B=0}=0
\eeq
so that the theory does not contain any intrinsic scale in its symmetric phase
$\langle\Phi_1\rangle=\langle\Phi_2\rangle=0$. However, on the lattice, due to
the relatively large errors introduced from the direct use of Eq.(20),
it is more convenient to define the massless theory as in \cite{ago} for the
one-component theory.
There, we started from the general expression \cite{zeit}
\beq
J(\phi_B) = \alpha \phi_B^3 \ln(\phi_B^2/v_B^2) +
\beta v^2_B \phi_B (1 - \phi_B^2/v_B^2),
\eeq
 which is still consistent with ``triviality'' (corresponding to an effective
potential given by the sum of a classical background and the zero point
energy of a free field) but allows for an explicit
scale-breaking term $\beta$. Setting
$\alpha=0$ one obtains a good description of the data in the ``extreme double
well'' limit ($r_o$ much more negative than $r_c$,
where $r_c$ corresponds to the onset of SSB)
where SSB is a semi-classical phenomenon and the
zero-point energy represents
a small perturbation. Then, we started to increase $r_o$, at
fixed $\lambda_o$, toward the unknown value $r_c$ and examined
the quality of the fit with $(\alpha,\beta,v_B)$ as free parameters. The
value of $r_o$
 at which the quality
of the 2- parameter fit $(\alpha,\beta=0,v_B)$ becomes exactly the same as that
of the
more general 3- parameter case was used to define the massless case.
 At $\lambda_o=1$
the massless regime of the one-component theory was found at a
 value  $r_o=r_s$ where $r_sa^2 \sim -0.45$. Finally,
 the accurate weak-coupling
relation between the bare mass and the euclidean cutoff \cite{cw}
($\lambda_B=6\lambda_o$)
\beq
 r_s=
-{{\lambda_B}\over{32\pi^2}}\Lambda^2
=-{{3\lambda_o}\over{16\pi^2}}\Lambda^2
\eeq
was used to generate the massless theory at different values of $\lambda_o$
from the previously determined
 value of $r_s$ at $\lambda_o=1$. It should be noted that our
numerical coefficient relating ultraviolet cutoff and lattice spacing for
quadratic divergences
($\Lambda a\sim 4.87$ ) agrees well with an independent analysis
of lattice data presented by Brahm \cite{bra} which gives
(in the range $\lambda_o \leq 10$ )
$\Lambda a=4.893\pm0.003$.
 Also, ref.\cite{bra} predicts the massless regime to correspond
to $r_sa^2=-(0.224\pm0.001)$ for
$\lambda_o=0.5$ in the infinite-volume limit. This implies
$r_sa^2\sim-(0.448\pm0.002)$ for $\lambda_o=1$,
in excellent agreement with our result.
\par In the O(2) case, one can use
  the results of \cite{ago,bra} by modifying Eq.(22)
for the two-component case. This simply introduces a combinatorial factor of
$4/3$ so that Eq.(22) becomes
\beq
 r_s=-{{\lambda_o}\over{4\pi^2}}\Lambda^2
\eeq
Hence, we expect the massless case to correspond to $r_sa^2\sim -0.6$ for
$\lambda_o=1$. This was confirmed
 to good accuracy by using the above-described fitting procedure to Eq.(21)
(now with $\phi_B$ replaced by $\rho_B$). Thus, the identification of the
massless regime on the lattice does seem to obey the simple scaling laws
(22-23) and is under theoretical control.
\par In the analysis of the
one-component theory it was found \cite{ago} that Eq.(3) was poorly
reproduced numerically at small values of $J$ ($a^3|J|\sim0.01$ or smaller).
As a consequence,
the values of $\phi_B$ (and any higher-order Green's functions)
extracted from the direct computation at $J= 0$ were not reliable. Similarly,
in the O(2) case we find that at small $J$
the exact $\theta$-independence
of $\rho_B$ (see Eq.(19) ),
is poorly reproduced and our data processing becomes unreliable.
We therefore consider a ``safe'' region
of $J$-values, $Ja^3 \geq 0.05$, in which the spurious $\theta$-dependence
is less than
$\pm 3\%$. Fitting the data to Eq.(15) we can infer the values
 of $v_B$ and $Z_{\phi}$ and compare with Eqs.(6,10).
\par Our numerical values for $a\rho_B(J)$ in the massless case,
 obtained with the Metropolis
algorithm on a $10^4$ lattice,
 are reported in Table I for
$\lambda_o=$1.0, 1.5 and 2.0.
For each $\lambda_o$ the corresponding $r_o$ is computed by using
Eq.(23) and our numerical input $r_sa^2=-0.6$ for $\lambda_o$=1 .
The errors in Table I
are essentially determined from the observed spurious variation of $\rho_B$
in the range $0\leq \theta \leq 2\pi$.
 As discussed above, this is a numerical
artifact and
 should be considered a systematic effect of the Monte Carlo lattice
simulation.
It is reproduced with three random number generators consistent
with the Kolmogorov-Smirnov test at the level O($10^{-4}$).
At low $J$ this systematic effect
completely dominates the error; the statistical errors, after 30,000
iterations, are 4-5 times smaller \cite{lange}.
\par Table I also reports the $v_B$ and $Z_{\phi}$ values obtained from the
two-parameter fits to the data using Eq.(15). The resulting $Z_{\phi}$ values
agree well with the one-loop prediction in Eq.(10).
To perform a more stringent test of the one-loop potential we next
constrain $Z_{\phi}$ to its one-loop value in Eq.(10)
and make a precise determination
of $av_B$ from a {\it one}-parameter
fit to Eq.(15). This allows a
meaningful comparison with the lattice version of Eq.(6)
\beq
  (av_B)^{\rm 1-loop}=
{{\pi y_{L}}\over{\sqrt{3\lambda_o }}} \exp(-{{8\pi^2}\over{9\lambda_o}})
\eeq
where we have identified the euclidean ultraviolet cutoff
$\Lambda \to {{\pi y_L}\over{a}}$, $y_L$ being an {\it a priori}
unknown coefficient. By replacing Eq.(24) into Eq.(15) we
 determined $y_L$ from the one-parameter fit to the data at
$\lambda_o=1.5$ and $r_oa^2=-0.9$ to be $y_L=2.44\pm0.03$
(${{\chi^2}\over{d.o.f.}}={{3.3}\over{17}}$). (As discussed in \cite{ago}, one
does not expect precisely the same numerical coefficient to govern the
relation between euclidean cutoff and lattice spacing for both quadratic and
logarithmic divergences, see below).
 In Table II we show
the results of the one-parameter fits to the data at $\lambda_o=$1.0 and 2.0
and the comparison with
Eq.(24) for $y_L=2.44\pm0.03$.
 It is apparent from Table II
that the one-loop potential well reproduces
the lattice data, as previously discovered in the one-component case
\cite{ago}.
\par The value of $y_L$ obtained from the
lattice simulation of the O(2) massless $\lambda\Phi^4$ theory is $\sim$17$\%$
larger than the value $y_L=2.07\pm0.01$ obtained in the one-component case
\cite{ago}. This is due to our choice of fixing the value of the bare
mass $r_sa^2=-0.6$ at $\lambda_o=1$ with the simple combinatorial
factor $4/3$ discussed above and ignoring finite size
corrections to Eq.(23), see \cite{bra}.
 Also, in the O(2) case, errors (as estimated from the spurious
$\theta$-dependence of $\rho_B$)
 are larger than in the
one-component theory and, moreover,
the massless
regime appears not so sharply identified on the basis of the $(J,\rho_B)$
correlation. Indeed, it is found in
a narrow range of $r_o$-values
near $r_oa^2=-0.6$ for $\lambda_o=1$. Choosing instead
 $r_sa^2\sim-0.58$ would give $y_L\sim 2.07$.
\par The agreement between one-loop predictions and Monte Carlo data is not
a trivial test of perturbation theory but, rather, represents a non
perturbative test of ``triviality''. If perturbation theory were valid then
the data should agree at least as well, if not better, with the
leading-log formula based on the perturbative $\beta$-function
\[
 J^{{\rm LL}}(\rho_B)={ {\lambda_o\rho^3_B}\over{1+{{5\lambda_o}
\over{4\pi^2}}
\ln{ { \pi x_{{\rm LL}} }\over{a\rho_B}} }}
\]
($x_{{\rm LL}}$ denoting an adjustable parameter). However, when we fit
the $\lambda_o=$ 1.0, 1.5 and 2.0 data to this formula
we find, respectively,
for 17 degrees
of freedom,
 $(\chi^2)_{{\rm LL}}=$9, 44 and 133 (to
compare with the values $\chi^2=$0.8, 3.3 and 7.3 obtained from the 1-loop
one-parameter fits
 with Eq.(15) when $Z_{\phi}$ is constrained to its value in Eq.(10)).
Note that we are in a region where the ``$\lambda_o$log'' term is not
small; it is of order unity. Thus, the good agreement between the data and
the one-loop formula is not because the higher-order corrections, expected
by perturbation theory, are negligibly small; it is because they are
{\it absent}. Without
the ``triviality'' argument, this would be an incomprehensible miracle.
\par In conclusion, the good agreement between lattice simulation and eqs.
(15,10,23,24)
 provides definite evidence that the dependence
of the effective potential on the radial field
in the continuous symmetry case is completely consistent with our expectations
\cite{consteve,zeit}.
This confirms that Eq.(11), up to small radiative corrections due to the gauge
and Yukawa couplings,
controls the relation between the Higgs mass and
the Fermi constant in the standard model if SSB is generated through
``dimensional
transmutation'' from a classically scale-invariant $\lambda\Phi^4$ theory.
 The ``triviality'' structure we have
checked, in which $m_h$ is not proportional to ``$\lambda_R$'' (which
vanishes),
implies that
the Higgs, despite of its rather large mass,
 is only weakly interacting and would be free if the gauge and Yukawa
couplings would be turned off.
\vskip 30 pt
\centerline{AKNOWLEDGEMENTS}
\par We thank P. M. Stevenson for very useful discussions and
collaboration.

\vfill
\eject
\widetext
\begin{table}
\caption
{ The values of $a\rho_B(J)$ for the massless case
are reported as discussed in the text. At the various values of $\lambda_o$
and $r_o$
we also show the results of the 2-parameter
fits with Eq.(15) and the one loop prediction (10).}
\label{Table I}
\end{table}
\begin{tabular}{lccc}
{}~~~$Ja^3$ &~~~~ {$\lambda_o=1.0~r_oa^2=-0.6$}~~~ &~~~
 {$\lambda_o=1.5~r_oa^2=-0.9$}
{}~~~ & ~~~
{$\lambda_o=2.0~r_oa^2=-1.2$} \\
\tableline
0.050 &$0.4110\pm0.0130$ & $0.3850\pm 0.0116$
&$0.3753\pm0.0091$ \\
\tableline
0.075 &$0.4686\pm0.0086$ & $0.4337\pm 0.0086$
&$0.4176\pm0.0069$ \\
\tableline
0.100&$0.5133\pm0.0084$ & $0.4724\pm 0.0065$
&$0.4517\pm0.0055$ \\
\tableline
0.125 &$0.5495\pm0.0063$ & $0.5048\pm0.0065$
&$0.4814\pm0.0054$ \\
\tableline
0.150&$0.5819\pm0.0060$ & $0.5332\pm 0.0053$
&$0.5063\pm0.0042$ \\
\tableline
0.200 &$0.6377\pm0.0047$ & $0.5812\pm 0.0040$
&$0.5500\pm0.0037$ \\
\tableline
0.250 &$0.6844\pm0.0039$ &$0.6217\pm0.0034$
&$0.5875\pm0.0029$\\
\tableline
0.300 &$0.7256\pm0.0030$ &$0.6571\pm0.0026$
&$0.6190\pm0.0029$ \\
\tableline
0.350 &$0.7612\pm0.0029$ &$0.6892\pm0.0026$
&$0.6473\pm0.0026$\\
\tableline
0.400 &$0.7938\pm0.0025$ & $0.7178\pm0.0025$
&$0.6731\pm0.0025$ \\
\tableline
0.450 &$0.8246\pm0.0024$ &$0.7435\pm0.0025$
&$0.6969\pm0.0023$\\
\tableline
0.500 &$0.8520\pm0.0023$ &$0.7683\pm0.0021$
&$0.7193\pm0.0023$ \\
\tableline
0.550 &$0.8785\pm0.0021$ &$0.7911\pm0.0021$
&$0.7399\pm0.0018$\\
\tableline
0.600 &$0.9029\pm0.0019$ &$0.8124\pm0.0020$
&$0.7593\pm0.0017$ \\
\tableline
0.650 &$0.9261\pm0.0019$ &$0.8330\pm0.0019$
&$0.7778\pm0.0017$\\
\tableline
0.700 &$0.9481\pm0.0018$ &$0.8520\pm0.0019$
&$0.7953\pm0.0017$\\
\tableline
0.750 &$0.9693\pm0.0018$ &$0.8703\pm0.0017$
&$0.8119\pm0.0015$\\
\tableline
0.800 &$0.9893\pm0.0016$ &$0.8879\pm0.0015$
&$0.8274\pm0.0013$\\
\tableline
{}~~~~~~&$Z_{\phi}=25.0\pm1.6$  &$Z_{\phi}=16.4\pm0.7$
&$Z_{\phi}=12.3\pm0.4$\\
\tableline
{}~~~~~~&$av_B=(1.4^{+1.5}_{-0.9})10^{-3}$  &$av_B=(1.8\pm 0.6)10^{-2}$
&$av_B=(5.6\pm0.8)10^{-2}$\\
\tableline
{}~~~~~~&$Z^{\rm 1-loop}_{\phi}=26.3$ &$Z^{\rm 1-loop}_{\phi}=17.5$
&$Z^{\rm 1-loop}_{\phi}=13.1$\\
\end{tabular}
\vfill
\eject
\begin{table}
\caption{By using Eq.(15), we show the results of the 1-parameter fits
for $av_B$ at $\lambda_o=$1.0 and 2.0 when
$Z_{\phi}$ is constrained to its one-loop value in Eq.(10). We also show the
predictions from Eq.(24), $(av_B)^{\rm Th}$, for $y_L=2.44\pm0.03$ as
determined from the fit to the data at $\lambda_o=$1.5 .}
\label{Table II}
\end{table}

\begin{tabular}{cc}
\tableline
{}~~~~~~~~~~~~~~~~{$\lambda_o=1.0~~r_oa^2=-0.6$}~~~~~~~~~~~~~~~~~~~~~ &
{$\lambda_o=2.0~~r_oa^2=-1.2$} \\
\tableline
$Z_{\phi}=26.32=fixed$ &$Z_{\phi}=13.16=fixed$\\
\tableline
$av_B=(7.09\pm0.11)10^{-4}$ &$av_B=(3.79\pm0.03)10^{-2}$\\
\tableline
${{\chi^2}\over{d.o.f}}={{0.8}\over{17}}$&${{\chi^2}\over{d.o.f}}=
{{7.3}\over{17}}$\\
\tableline
$(av_B)^{\rm Th}=(6.85\pm0.09)10^{-4}$ &
$(av_B)^{\rm Th}=(3.89\pm0.05)10^{-2}$
\end{tabular}


\begin{references}
  \bibitem{huang}
K. Huang, E. Manousakis, and J. Polonyi, Phys. Rev. {\bf D35}, 3187 (1987);
K. Huang, Int. J. Mod. Phys. {\bf A4}, 1037 (1989); in Proceedings of the
DPF Meeting, Storrs, CT, 1988.
  \bibitem{ago}
A. Agodi, G. Andronico, and M. Consoli, {\it The real test of ``triviality''
on the lattice}, INFN and University of Catania preprint, February 1994,
submitted to Physical Review Letters (hep-th-9402071).
  \bibitem{consteve}
M. Consoli and P. M. Stevenson, {\it Resolution of the $\lambda\Phi^4$
Puzzle and a 2 TeV Higgs Boson}, Rice University preprint,
DE-FG05-92ER40717-5, July 1993, submitted to Physical Review D. (hep-ph
9303256).
  \bibitem{zeit}
M. Consoli and P. M. Stevenson, {\it The Non-Trivial Effective Potential
of the ``Trivial'' $\lambda\Phi^4$ Theory: A Lattice Test}
, Rice University preprint,
DE-FG05-92ER40717-9, October 1993, to appear in Zeit.f\"ur Physik C.
  \bibitem{froh}
J. Fr\"ohlich, Nucl. Phys. {\bf B200}(FS4), 281 (1982);
R. Fern\'andez, J. Fr\"ohlich, and A. D. Sokal, {\it Random Walks, Critical
Phenomena, and Triviality in Quantum Field Theory} (Springer-Verlag, Berlin,
1992).
  \bibitem{trivpert}
M. Consoli and P. M. Stevenson, {\it ``Triviality''and the perturbative
expansion in $\lambda\Phi^4$ theory}, Rice University preprint, March 1994,
submitted to Physics Letters B.
  \bibitem{bran}
V. Branchina, P. Castorina, M. Consoli and D. Zappala', Phys. Lett.
{\bf B274}, 404 (1992).
  \bibitem{con}
M. Consoli, in ``Gauge Theories Past and Future -- in Commemoration
of the 60th birthday of M. Veltman'', R. Akhoury, B. de Wit,
P. van Nieuwenhuizen and H. Veltman Eds., World Scientific 1992, p. 81;
M. Consoli, Phys. Lett. {\bf B 305}, 78 (1993).
  \bibitem{iban}
R. Iba\~nez-Meier and P. M. Stevenson, Phys. Lett. {\bf B297}, 144 (1992).
  \bibitem{new}
V. Branchina, M. Consoli and N. M. Stivala, Zeit. Phys.{\bf C57}, 251 (1993).
  \bibitem{rit}
U. Ritschel, Phys. Lett. {\bf B318}, 617 (1993).
  \bibitem{cian}
M. Consoli and A. Ciancitto, Nucl. Phys. {\bf B254}, 653 (1985).
  \bibitem{return}
P. M. Stevenson and R. Tarrach, Phys. Lett. {\bf B176}, 436 (1986).
  \bibitem{cast}
P. Castorina and M. Consoli, Phys. Lett. {\bf B235}, 302 (1990); V. Branchina,
P. Castorina, M. Consoli and D. Zappala', Phys. Rev. {\bf D42}, 3587 (1990).
  \bibitem{dj}
L. Dolan and R. Jackiw, Phys. Rev. {\bf D9}, 3320, (1974).
  \bibitem{cw}
S. Coleman and E. Weinberg, Phys. Rev. {\bf D7}, 1888 (1973).
  \bibitem{bra}
D. E. Brahm, {\it The lattice cutoff for $\lambda\Phi^4_4$ and
$\lambda\Phi^6_3$}, CMU-HEP94-10 preprint, hep-lat/9403021, March 1994.
  \bibitem{lange}
We have found, however, that the
use of the Langevin formulation of the lattice theory considerably reduces this
unphysical effect thus allowing, in principle,
a more precise determination of the $(J,\rho_B)$ correlation.
We are presently investigating  this alternative approach.
\end{references}
\end{document}